# Using Model-Assisted Calibration Methods to Improve Efficiency of Regression Analyses with Two-Phase Samples under Complex Survey Designs

Lingxiao Wang

Department of Statistics, University of Virginia, U.S.A.

National Cancer Institute, Division of Cancer Epidemiology & Genetics, Biostatistics Branch, U.S.A.

Address: Department of Statistics, University of Virginia, Charlottesville, VA 22903.

Email: lingxiao.wang@virginia.edu

**SUMMARY**（203 words）

Two-phase sampling designs are frequently employed in epidemiological studies and large-scale health surveys. In such designs, certain variables are exclusively collected within a second-phase random subsample of the initial first-phase sample, often due to factors such as high costs, response burden, or constraints on data collection or measurement assessment. Consequently, second-phase sample estimators can be inefficient due to the diminished sample size. Model-assisted calibration methods have been used to improve the efficiency of second-phase estimators. However, no existing methods provide appropriate calibration auxiliary variables while simultaneously considering the complex sample designs present in both the first- and second-phase samples in regression analyses. This paper proposes to calibrate the sample weights for the second-phase subsample to the weighted entire first-phase sample based on score functions of regression coefficients by using predictions of the covariate of interest, which can be computed for the entire first-phase sample. We establish the consistency of the proposed calibration estimation and provide variance estimation. Empirical evidence underscores the robustness of the calibration on score functions compared to the imputation method, which can be sensitive to misspecified prediction models for the variable only collected in the second phase. Examples using data from the National Health and Nutrition Examination Survey are provided.

# 1 INTRODUCTION

For studies with two-phase sample designs, a random subsample is selected from the initial first-phase sample to collect further measurements from only the subsample when these measurements are often costly or otherwise not possible to collect in the first-phase sample. Various types of two-phase sample designs are available (Tao et al., 2020) where some are typically used in epidemiologic studies. The first-phase sample is usually an existing cohort that is treated as simple random sample from an infinite superpopulation. A subsample is then randomly selected in the second phase using various sample designs, e.g., case-cohort or nested case-control studies (Prentice, 1986; Breslow and Hu, 2018). Large-scale epidemiologic studies, such as population-based cohort and case-control studies may also use complex sample designs including stratified multistage cluster sampling in selecting first-phase samples to better represent the target population (Colt et al., 2011; Tota et al., 2019; LaVange et al., 2010). However, the existing analyses conducted in two-phase epidemiologic studies may or may not account for first-phase complex sampling, e.g., the Hispanic Community Health Study/Study of Latinos (Chen et al., 2020; Goodman et al., 2021).

Two-phase sampling is also commonly used in national health surveys, such as the US National Health and Nutrition Examination Survey (NHANES) and the US National Health Interview Survey, to reduce costs and respondent burden (Neyman, 1938; Thompson, 1992 Chapter 14). In contrast to epidemiologic studies, complex survey analyses consider sampling designs for both first- and second-phase samples in national health surveys. For example, NHANES selects representative random samples of individuals from a target finite population ($FP$) of non-institutionalized US residence by stratified multistage cluster sample designs. In the first stage, primary sampling units (PSUs) of counties (or groups of contiguous counties) are selected with probability proportional to measures of size (PPS) within each stratum defined by geography (e.g., census regions). In later stages, segments of census blocks, households, and individuals are randomly selected within each sampling unit sampled in the early stage sequentially with unequal selection probabilities. This initial stratified multistage cluster sample selected from the $FP$ is treated as the first-phase full survey sample where abundant data on diseases/conditions and risk factors are collected through questionnaires, physical examinations, and laboratory tests of biospecimens. Subsamples of the first-phase survey sample are then randomly selected in the second phase to collect an additional covariate (or covariates) of interest that can be important predictors of the disease outcome (the type I two-phase design in Table 1). In regression analyses, complex sample designs in either of the phases should be accounted for so that consistent estimators of $FP$ quantities can be obtained.

Constraining the analysis to the complete second-phase sample may reduce the efficiency due to the variable sample weights and decreased sample sizes for both national health surveys and population-based epidemiologic studies (LaVange, 2010). In national surveys, the classical calibration methods (e.g., poststratification) calibrate the sample weights of both first-phase and second-phase samples to the $FP$ on covariates such as age, race/ethnicity and sex known for the $FP$. Calibration weighting on covariates have been shown to improve efficiency of the sample weighted total/mean/proportion estimators when the covariates are highly correlated with the outcomes of interest (Deville & Sarndal, 1992).

However, the classic calibration to the $FP$ does not make full use of variables that are collected in the first-phase complex survey sample. Beaumont et al. (2015) calibrated second-phase sample weights to the weighted first-phase sample on covariates collected in the first phase. Nevertheless, it is known from the two-phase literature that calibrating on covariates cannot improve efficiency of regression analyses, because the covariates are not correlated with regression coefficient estimators, even though they can be strongly associated with the outcome. (Lumley et al., 2011).

To utilize the variables collected in the first-phase sample, including the outcome, covariates, along with additional "ancillary" variables (i.e., good predictors for the second-phase covariate of interest, but uncorrelated with the outcome conditional on all covariates), this paper proposes to calibrate the second-phase sample weights to the weighed first-phase sample on score functions of the regression coefficients that use predictions for covariates of interest that are only available in the second phase sample. The predictions are obtained for the entire first-phase sample using the ancillary variables as predictors.

Besides the typical two-phase design (type I), another two-phase design (type II) can occur in repeated cross-sectional health surveys when some variables are measured in only certain survey cycles. For example, Liver Ultrasound Elastography was conducted in NHANES 2017-2018. As NHANES is series of continuous biannual cycles of national surveys, combining samples from multiple cycles is a common way to increase sample size for efficiency improvement in survey research (Korn and Graubard, 1999; Heeringa et al., 2010). In this paper, we innovatively consider combining samples from multiple cycles as a two-phase complex design where samples from all combined cycles and from the cycles that measure more variables are respectively treated as the "first" and the "second phase" samples (Table 1). Section 2 discusses the assumptions needed for the type II two-phase sample design to produce approximately unbiased estimation. Under both two-phase designs, we consider the data structure for regression analyses where an important covariate of interest is only collected in the second phase while the outcome of interest, other covariates in the regression model, and some extra ancillary variables are collected in the first phase for illustration purpose (Table 1). The method can be easily applied to data structures where multiple covariates, or the outcome are only collected in the second phase sample as well.

The major contribution of this paper is to provide a general calibration weighting using score functions to substantially improve efficiency of regression analyses under the two types of two-phase designs nested in complex sample designs. The proposed inference with calibration weighting is different from the weight calibration on influence functions in the typical cohort studies (Breslow and Hu, 2018; Lumley et al., 2011; Shin et al., 2020) where the first-phase cohort sample is treated as a simple random sample of an infinite superpopulation (Samuelsen, 1997; Shin et al., 2020). Those approaches can be difficult to apply under two-phase designs in national surveys or population-based epidemiologic studies, where both complex sample designs for both first- and second-phase sample, and the correlation between the two samples need to be considered in inferences of regression analyses utilizing weight calibration. The proposed calibration approach not only simplifies the calibration weighting by using score functions instead of influence functions as the calibration auxiliary variables, but also considers the complex sampling designs in both phases.

We prove that the proposed model-assisted calibration estimators are consistent to the finite population quantities and provides consistent Taylor linearization variance estimators. Simulations are used to study the finite sample performance of the calibration method compared with the traditional survey calibration method on covariates, as well as the imputation method under correctly specified and misspecified imputation models. The calibration method is applied to estimating odds ratios of two-hour fasting glucose tolerance for all-cause mortality, and of Non-Alcoholic Fatty Liver Disease (NAFLD) for diabetes in the US under type I and type II two-phase sample designs nested in NHANES, respectively.

## 2 NOTATION AND MODEL

We illustrate the method using logistic regression, but the method can be easily applied to other analyses such as multiple linear regression and Cox regression. We aim to estimate the risk of developing an outcome, e.g., disease, $Y$ (1 for presence, and 0 for absence of disease) given a set of covariates $X = (X_1, X_2)^T$, i.e., $r(X) = \Pr(Y = 1 \mid X)$, with special focus on the association between the covariate $X_2$ with $Y$ in the target finite population ($FP$) with $N$ individuals indexed by $i \in \{1, \cdots, N\}$. For simplicity, we assume $X_2$ is univariate, but the method can be easily applied to multivariate $X_2$. We posit a logistic regression model in the $FP$,

$$\text{logit}\{\Pr(Y = 1 \mid X)\} = \beta_0 + \boldsymbol{\beta}^T X_1 + \beta_2 X_2, \tag{2.1}$$

where $\beta_0$ and $\boldsymbol{\beta} = (\boldsymbol{\beta}_1, \beta_2)^T$ are the log-odds ratios (log-OR's). Besides $X$ and $Y$, there are a set of "ancillary" variables $Z$ that can be highly predictive of $X_2$ but independent of $Y$ conditional on $X$. The values of $X$, $Y$, and $Z$ are only observed in random samples of $FP$. We consider two types of two-phase sample designs nested within a complex sample design for observation of $X$, $Y$, and $Z$ described Section 1 and Table 1. Details are as follows.

**Type I Two-Phase Sample Design** Let every individual in the $FP$ have a positive probability of being randomly included in a first-phase sample $s_1$ (with sample size $n_1$), denoted by $\{\pi_{1,i} > 0, i \in FP\}$. We use $\delta_{1,i}$ to indicate if individual $i \in FP$ is selected in $s_1$ ($\delta_{1,i} = 1$ if $i \in s_1$, and $\delta_{1,i} = 0$ if $i \in FP - s_1$). The sample weight for $i \in s_1$ is $w_{1,i} = \pi_{1,i}^{-1}$. The observed values of $X_1$, $Y$, and $Z$ are denoted by $(x_{1,i}, y_i, z_i)$ for $i \in s_1$. The second-phase sample, $s_2 \subset s_1$ is a random subsample of $s_1$ (with sample size $n_2 < n_1$), where the second-phase inclusion probabilities from $s_1$ are $\{\pi_{2,i} > 0, i \in s_1\}$, with the observed weights, $\{w_{2i} = \pi_{2,i}^{-1}, i \in s_2\}$ that weight $s_2$ up to $s_1$. Values of $X_2$, denoted by $x_{2,i}$, are observed for $i \in s_2$ only. The inclusion probabilities from $FP$ to $s_2$ for $FP$ individuals and the observed sample weights for $s_2$ individuals across the two phases are $\{\pi_i = \pi_{1,i}\pi_{2,i}, i \in FP\}$ and $\{w_i = w_{1,i}w_{2,i}, i \in s_2\}$, respectively.

**Type II Two-Phase Sample Design** Let there be $C$ survey samples, denoted by $s^{(1)}, \cdots, s^{(C)}$ which are randomly selected from the $FP$ in survey cycles $1, \cdots, C$, respectively. We observe $(x_{1,i}, y_i, z_i)$ in all $C$ survey samples. The cycles are typically at regular time intervals (e.g., annually, or biannually). The "first-phase sample" is the union $s_1 = s^{(1)} \cup \cdots \cup s^{(C)}$. Let $\pi_i^{(c)}$ be the probability that $i \in FP$ is included in $s^{(c)}$, with the corresponding sample weight $w_i^{(c)} = \{\pi_i^{(c)}\}^{-1}, c = 1, \cdots, C$. We make the following standard assumptions for combining cycles of a survey over time (Heeringa et al., 2010):

(A.1) The target $FP$ is approximately stationary, over the time of $C$ survey cycles.

(A.2) The range of sample weights does not change much over the $C$ survey cycles.

(A.3) The $C$ samples are independently selected from the $FP$ across the $C$ survey cycles.

Under (A.1), the "first-phase sample" $s_1$ is treated as a random sample of the $FP$. Under (A.2), the probability of being included in $s_1$ from the $FP$ is calculated by $\pi_{1,i} = C\pi_i^{(c)}$ with the sample weight $w_{1,i} = \{C\pi_i^{(c)}\}^{-1}$ for $i \in s^{(c)}$ (NHANES, 2018). Assumption (A.3) is needed for variance calculation and efficiency gain evaluation (see Section 3.2 and more discussion in Section 6).

The "second-phase sample" $s_2 = s^{(1)} \cup \cdots \cup s^{(B)}$ is the union of a subset of samples from $B < C$ survey cycles where values of $X_2$ are observed. The inclusion probability for $i \in s_2$ "sampled" from $s_1$ is $\pi_{2,i} \equiv B/C$ under assumptions (A.1) and (A.2). The $FP$ inclusion probabilities and sample weights for $i \in s_2$ across the "two phases of sampling" are respectively $\pi_i = \pi_{1,i}\pi_{2,i} = B\pi_i^{(k)}$ and $w_i = w_{1,i}w_{2,i} = \{B\pi_i^{(k)}\}^{-1}$ for $i \in s^{(k)}, k = 1, \dots, B$ (NHANES, 2018).

## 3 THE PROPOSED MODEL-ASSISTED CALIBRATION WEIGHTING METHOD

Calibration weighting (Deville and Sarndal, 1992) has been widely used in survey research to reduce bias in Horvitz-Thompson (HT) estimators of $FP$ total and mean of the outcome $Y$ due to under coverage (e.g., not including institutional subpopulations such as military personnel or nursing home residents in the survey sampling frame) by borrowing strength from summary statistics of "calibration auxiliary variables" $\boldsymbol{v}$ that are available in the $FP$ and correlated with the outcome $Y$. In traditional calibration weighting methods, $\boldsymbol{v}$ only include a limited set of covariates such as demographic variables (age/sex/race). Because $FP$ summary statistics of $\boldsymbol{v}$ can be treated as having small-order variability, the calibrated estimation is also more efficient than the HT estimation when $\boldsymbol{v}$ and $Y$ are highly correlated (Deville & Sarndal, 1992). This constraint limits their application to more complex analyses, such as estimating log-OR's $\boldsymbol{\beta}$ in model (2.1) due to the weak correlation between $\boldsymbol{v}$ and the $\boldsymbol{\beta}$ estimator, despite the strong association of $\boldsymbol{v}$ and $Y$.

To improve efficiency of regression analyses under two-phase complex survey designs, we propose to calibrate the sample weights of $s_2$ to the sample weighted $s_1$ using auxiliary variables $\boldsymbol{v}$ generated from score equations for estimating log-OR's, which fully uses the information of $Y$, $\boldsymbol{X}_1$, and $Z$ observed in $s_1$. In $FP$ inferences, calibrating on score functions can be very useful, especially with the commonly employed type I and type II two-phase designs in national surveys.

### 3.1 Calibration Estimator of Log-OR's Using Score Functions as Auxiliary Variables

The calibration weights for $s_2$, denoted by $\{\widetilde{w}_i, i \in s_2\}$, is obtained by minimizing the averaged distance between $\{\widetilde{w}_i, i \in s_2\}$ and the original sample weights $\{w_i, i \in s_2\}$ with the constraint $\sum_{i \in s_2} \widetilde{w}_i \boldsymbol{v}_i = \sum_{i \in s_1} w_{1,i} \boldsymbol{v}_i$, This is equivalent to solving the following equation:

$$Q(\boldsymbol{\eta}) = \frac{1}{N}\left\{\sum_{i\in s_2} F(\boldsymbol{v}_i^T\boldsymbol{\eta})w_i\boldsymbol{v}_i - \sum_{i\in s_1} w_{1,i}\boldsymbol{v}_i\right\} = \boldsymbol{0}, \tag{3.1}$$

where $F(\cdot)$ is the calibration factor determined by the selected distance function (Deville and Sarndal, 1992), $\boldsymbol{v}$ is a vector of calibration auxiliary variables, and $\boldsymbol{\eta}$ is the nuisance parameter. For example, when the chi-squared distance (Deville and Sarndal, 1992) is used, we have $F(\boldsymbol{v}_i^T\boldsymbol{\eta}) = \boldsymbol{v}_i^T\boldsymbol{\eta} + 1$. The solution of equation (3.1), denoted by $\widehat{\boldsymbol{\eta}}$, has the following explicit form:

$$\widehat{\boldsymbol{\eta}} = \left(\sum_{i\in s_2} w_i \boldsymbol{v}_i^T \boldsymbol{v}_i\right)^{-1} \left(\sum_{i\in s_1} w_{1,i}\boldsymbol{v}_i - \sum_{i\in s_2} w_i\boldsymbol{v}_i\right) \tag{3.2}$$

The resulting calibration weights are $\{\widetilde{w}_i = F(\boldsymbol{v}_i^T\widehat{\boldsymbol{\eta}})w_i, i \in s_2\}$.

The auxiliary variables $\boldsymbol{v}$ should be available in the entire $s_1$ and highly correlated with the sample estimator of $\boldsymbol{\beta}$. If the covariate of interest $X_2$ is available in the entire $s_1$, we can estimate $\boldsymbol{\beta}$ from $s_1$ by solving the score equation:

$$\boldsymbol{U}(\boldsymbol{\beta}) = \sum_{i\in FP} \delta_{1,i} w_{1,i}(y_i - p_i)(1, \boldsymbol{x}_i^T)^T = \boldsymbol{0} \tag{3.3}$$

where $\boldsymbol{x}_i = (\boldsymbol{x}_{1,i}^T, x_{2,i})^T$, $p_i = \text{expit}\{(1, \boldsymbol{x}_i^T)\boldsymbol{\beta}\}$. The resulting estimator, denoted by $\widehat{\boldsymbol{\beta}}^{(s_1)}$, is consistent to $\boldsymbol{\beta}$ and more efficient than the $s_2$ estimator because of the larger sample size $n_1 > n_2$. In Supplementary Materials A.1, A.2, it is shown that

$$\widehat{\boldsymbol{\beta}}^{(s_1)} - \boldsymbol{\beta} = \sum_{i\in s_1} w_{1,i}\Delta_i\{\widehat{\boldsymbol{\beta}}^{(s_1)}\} + o\{n_1^{-1/2}\}, \tag{3.4}$$

where $\Delta_i\{\widehat{\boldsymbol{\beta}}^{(s_1)}\} = N^{-1}\boldsymbol{U}_{\boldsymbol{\beta}}^{-1}(y_i - p_i)(1, \boldsymbol{x}_i^T)^T$ is the influence function of $\widehat{\boldsymbol{\beta}}^{(s_1)}$ and $\boldsymbol{U}_{\boldsymbol{\beta}} = \partial\boldsymbol{U}(\boldsymbol{\beta})/\partial\boldsymbol{\beta}$. Notice that $\Delta_i\{\widehat{\boldsymbol{\beta}}^{(s_1)}\} \propto \boldsymbol{u}_i(\boldsymbol{\beta})$ where $\boldsymbol{u}_i(\boldsymbol{\beta}) = (y_i - p_i)(1, \boldsymbol{x}_i^T)^T$ is the unweighted contribution of the $i \in s_1$ to the entire score function $\boldsymbol{U}(\boldsymbol{\beta})$. The sample plug-in individual score functions are obtained by

$$u_i\{\widehat{\boldsymbol{\beta}}^{(s_1)}\} = (y_i - p_i)(1, \boldsymbol{x}_i^T)^T\big|_{\boldsymbol{\beta}=\widehat{\boldsymbol{\beta}}^{(s_1)}}. \tag{3.5}$$

Formulae (3.4) and (3.5) implies that $\widehat{\boldsymbol{\beta}}^{(s_1)}$ can be highly correlated with the score functions $u_i\{\widehat{\boldsymbol{\beta}}^{(s_1)}\}$. Therefore, the calibration auxiliary variables $\boldsymbol{v}_i$ in equations (3.1) should include $\{u_i\{\widehat{\boldsymbol{\beta}}^{(s_1)}\}, i \in s_1\}$. However, because $X_2$ is **not** observed in $s_1 - s_2$, $\boldsymbol{\beta}$ cannot be directly estimated from $s_1$. Consequently, $u_i\{\widehat{\boldsymbol{\beta}}^{(s_1)}\}$ cannot be obtained for $s_1$. An alternative calibration variable $\boldsymbol{v}_i$ that are highly correlated with $u_i\{\widehat{\boldsymbol{\beta}}^{(s_1)}\}$ can be generated from variables measured in the entire $s_1$, i.e., $\{w_{1,i}, \boldsymbol{x}_{1,i}, y_i, \boldsymbol{z}_i, i \in s_1\}$. The procedure of obtaining substitutes of $u_i\{\widehat{\boldsymbol{\beta}}^{(s_1)}\}$ as auxiliary variables $\boldsymbol{v}_i$ and creating the calibration weights are described below and in Figure 1.

**Step 1** Fit a prediction model of $X_2$ in $s_2$ using variables $(\boldsymbol{X}_1^T, \boldsymbol{Z}^T)^T$ that are available in $s_1$ as predictors and predict $X_2$ for the entire $s_1$ using this prediction model. Use $X_2^*$ to denote the prediction variable for $X_2$.

**Step 2** Fit the regression model in $s_1$ by using the prediction $X_2^*$ to replace $X_2$ in Model (2.1). The corresponding regression coefficients $\boldsymbol{\beta}^*$ are estimated by solving the $s_1$ estimating equation:

$$U^*(\boldsymbol{\beta}^*) = \frac{1}{N}\sum_{i \in s_1} w_{1,i}(y_i - p_i^*)(1, \boldsymbol{x}_i^{*T})^T = \boldsymbol{0}, \tag{3.6}$$

where $\boldsymbol{x}_i^* = (\boldsymbol{x}_{1,i}^T, x_{2,i}^*)^T$, $x_{2,i}^*$ is the computed value of $X_2^*$ for $i \in s_1$ and $p_i^* = \text{expit}\{(1, \boldsymbol{x}_i^{*T})\boldsymbol{\beta}^*\}$. We denote the solution of equation (3.6) as $\widehat{\boldsymbol{\beta}}^*$, and obtain the plug-in score functions of $\widehat{\boldsymbol{\beta}}^*$:

$$u_i^*\{\widehat{\boldsymbol{\beta}}^*\} = (y_i - p_i^*)(1, \boldsymbol{x}_i^{*T})^T\big|_{\boldsymbol{\beta}^*=\widehat{\boldsymbol{\beta}}^*}, i \in s_1 \tag{3.7}$$

**Step 3** Use $\boldsymbol{v}_i = u_i^*\{\widehat{\boldsymbol{\beta}}^*\}$ as the calibration auxiliary variables in equation (3.1) to obtain the calibration adjustment factor $F_i = F(\boldsymbol{v}_i^T\widehat{\boldsymbol{\eta}})$ $i \in s_2$.

Finally, fit the outcome model (2.1) to the calibration weighted sample $s_2$. The calibration estimator, denoted by $\widehat{\boldsymbol{\beta}}^{(calib)}$, is obtained by solving the calibration weighted score equation:

$$\boldsymbol{U}(\boldsymbol{\beta}) = \frac{1}{N}\sum_{i \in s_2} F_i w_i (y_i - p_i)(1, \boldsymbol{x}_i^T)^T = \boldsymbol{0}, \qquad (3.8)$$

Under regular conditions C1-C6 in Supplementary Materials A.1, $\widehat{\boldsymbol{\beta}}^{(calib)}$ is a consistent estimator of $\boldsymbol{\beta}$ with $\widehat{\boldsymbol{\beta}}^{(calib)} = \boldsymbol{\beta} + O(n_2^{-1/2})$ (Supplementary Materials A.3).

Notice that estimating $\boldsymbol{\beta}$ via solving the calibration weighted estimating equation (3.8) that uses $\boldsymbol{u}_i^*\{\boldsymbol{\beta}^*\}$ as the calibration auxiliary variables is equivalent to estimating $\boldsymbol{\beta}$ via solving the following constrained estimating equation.

$$\boldsymbol{U}(\boldsymbol{\beta}) = \frac{1}{N}\sum_{i \in s_2} F_i w_i (y_i - p_i)(1, \boldsymbol{x}_i^T)^T = \boldsymbol{0} \text{ subject to}$$

$$F_i > 0, \text{ and } \frac{1}{N}\sum_{i \in s_2} F_i w_i (y_i - p_i^*)(1, \boldsymbol{x}_i^{*T})^T = \boldsymbol{U}^*(\boldsymbol{\beta}^*) = \boldsymbol{0}. \qquad (3.9)$$

## 3.2 Sample Variance Estimation for $\widehat{\boldsymbol{\beta}}^{(calib)}$

Under the Type I two-phase sample design, we assume a general case where $s_1$ is randomly selected from the $FP$ by a stratified multistage cluster design. The sample fraction of PSUs in each stratum is assumed to be small. We then suppose $s_2$ is randomly selected from $s_1$ using a stratification sample design where the PSU's in $s_1$ are treated as strata so that the number of strata and PSUs in $s_1$ and $s_2$ are respectively the same (NHANES 2018). Under the Type II two-phase sample design, $s^{(1)}, \cdots, s^{(C)}$ are also randomly selected from the $FP$ by a stratified multistage cluster design, and they are treated as strata (in addition to the sampling strata within each $s^{(1)}, \cdots, s^{(C)}$) in $s_1$ (NHANES 2018). Please refer to Supplementary Materials A.3 for the complete Taylor Linearization (TL) sample variance estimation for $\widehat{\boldsymbol{\beta}}^{(calib)}$ that considers variability due to sampling $s_1$ and $s_2$, and refer to Supplementary Materials B for a simplified TL variance estimation under the two types of two-phase sample designs when the Chi-square distance $G(u) = (u-1)^2/2$ is used for weight calibration.

## 4 SIMULATIONS

The purpose of the simulations is to help assess finite sample biasness and efficiency of the proposed calibration method in comparison to other approaches described in Section 4.3 when estimating log-OR's under the Type I and II two-phase complex sample designs. The simulation setup varies over the following three factors that can affect the efficiency gains of the calibration estimates compared to the direct weighted $s_2$ estimates: (1) the correlation between the covariate of interest $X_2$ and other covariates $X_1$; (2) the correlation between $X_2$ and its prediction; and (3) the sample fraction of the $s_2$ among $s_1$ (i.e., $f_2 = n_2/n_1$).

### 4.1 Finite population ($FP$) generation

An $FP$ of size $N = 2{,}000{,}000$ is generated with an outcome $Y$ that depends on three covariates $\boldsymbol{X}_1 = (X_{1,1}, X_{1,2})^T$, and $X_2$, where $X_2$ is the covariate of interest and it is generated from ancillary variables $\boldsymbol{Z} = (Z_1, Z_2)^T$. $Z_1 \sim N(0,1)$; $(X_{1,1}, Z_2)^T \sim N(\boldsymbol{0}, \boldsymbol{\Sigma})$, where $\boldsymbol{\Sigma} = \begin{pmatrix} 1 & \rho_{X_{1,1}Z_2} \\ \rho_{X_{1,1}Z_2} & 1 \end{pmatrix}$, $X_{1,2} \sim Bernoulli(0.3)$; and

$$X_2 = 1 + 0.5Z_1 + 1.5Z_2 + \epsilon, \tag{4.1}$$

with $\epsilon \sim N(0, 0.5)$. The empirical correlation between $X_1$ and $X_2$, which depends on $\rho_{X_{1,1}Z_2}$, is varied from 0.09 to 0.81 in the $FP$ to investigate the impact of factor (1).

Three predictions of $X_2$ are generated in the $FP$ to investigate the impact of factor (2): $X_2^* = 1 + 0.5Z_1 + 1.5Z_2$, $X_2^{**} = -1.9 + 0.5Z_1 + 1.45\tilde{Z}_2$, and $X_2^{***} = 1 + 0.5Z_1$ with $\tilde{Z}_2$ defined by $\tilde{Z}_2 = 1$ if $Z_2 \leq Q_{40}^{(Z_2)}$, $\tilde{Z}_2 = 2$ if $Q_{40}^{(Z_2)} < Z_2 \leq Q_{60}^{(Z_2)}$ and $\tilde{Z}_{2,i} = 3$ if $Z_{2,i} > Q_{60}^{(Z_2)}$ where $Q_u^{(Z_2)}$ is the $u^{\text{th}}$ percentile of $Z_2$. Notice that $X_2^*$ is generated using the true prediction model of $X_2$ in (4.1) while $X_2^{**}$ and $X_2^{***}$ are generated from misspecified prediction models of $X_2$. Therefore, the correlation between $X_2$ and the three predictors of $X_2$ differ, with the empirical values $\rho_{X_2,X_2^*} \approx 0.95$, $\rho_{X_2,X_2^{**}} \approx 0.84$, and $\rho_{X_2,X_2^{***}} \approx 0.30$ throughout the simulation. The outcome $Y$ is generated as $Y_i \sim Bernoulli(p_i)$, with $p_i = \text{logit}(\beta_0 + \boldsymbol{\beta}^T \boldsymbol{x}_i)$, where $\beta_0 = -3$, $\boldsymbol{\beta} = (\beta_{1,1}, \beta_{1,2}, \beta_2, \beta_{2,2})^T = (-3, 0.7, 0.9, 0.5, 0.3)^T$, and $\boldsymbol{x}_i = (x_{1,1,i}, x_{1,2,i}, x_{2,i}, x_{2,i} \cdot x_{1,2,i})^T$ are realizations of $\boldsymbol{X} = (\boldsymbol{X}_1^T, X_2, X_2 \cdot X_{1,2})^T$ for $i \in FP$. We are particular interested in $\beta_2$ and $\beta_{2,2}$, which are the effects of $X_2$ and its interaction with $X_{1,2}$ respectively.

### 4.2 Selecting Samples from the $FP$ under Two Types of Two-Phase Complex Designs

**Type I Design** The first-phase sample $s_1$ of $n_1 = 10{,}000$ individuals are randomly selected from $FP$ by a two-stage cluster sampling design with proportional-to-size (PPS) with replacement sampling at each stage. The $s_2$ is a random sample of $s_1$ with sample sizes of $n_2$ that varied from 1,000 to 5,000. Correspondingly, the second-phase sample fraction $f_2 = n_2/n_1$ varies from 0.1 to 0.5. to examine the impact of factor (3). See Supplementary Materials C for details.

**Type II Design** Let $s_1 = s^{(1)} \cup \cdots \cup s^{(C)}$ where $s^{(1)}, \cdots, s^{(C)}$ are independently randomly selected from the $FP$ with respective sample sizes $n^{(1)}, \cdots, n^{(C)}$. For simplicity, we set $n^{(1)} = \cdots = n^{(C)} = n$ and the total sample size of $s_1$ is $n_1 = Cn$. The same two-stage cluster PPS sample design for selecting $s_1$ under the Type I design is used to obtain each of the $s^{(1)}, \cdots, s^{(C)}$. Without loss of generality, we assume $s_2 = s^{(1)}$ with $n_2 = n$. To investigate the impact of factor (3), we vary $C$ from 10 to 2, so that $f_2 = n_2/n_1$ increases from 0.1 to 0.5, respectively. We fix $n_1 = 10{,}000$, then $n^{(1)} = \cdots = n^{(C)} = n = n_2$ is varied from 5,000 to 1,000 correspondingly, so that the sample sizes of $s_1$ and $s_2$ under the two types of designs are respectively comparable.

## 4.3 Simulation Results

The proposed calibration method that uses one of $X_2^*$, $X_2^{**}$, or $X_2^{***}$ as the prediction for $X_2$ are respectively denoted by Calib.$X_2^*$, Calib.$X_2^{**}$ and Calib.$X_2^{***}$. The proposed calibration estimators of log-OR's $\boldsymbol{\beta}$ are compared to the following 6 methods. (1) Direct.$s_2$: the original $s_2$ sample weighted estimators; (2) Calib.FP: the traditional calibration method in equation (3.1) that calibrates the weighted $s_2$ to $FP$ on covariates $\boldsymbol{X}_1$ and $\boldsymbol{Z}$; (3) Direct.$s_1$: $s_1$ sample weighted estimators, which is expected to be the most efficient among all considered methods, but not available in practice due to missing $X_2$ in $s_1 - s_2$; (4) Imp.$X_2^*$, (5) Imp.$X_2^{**}$, and (6) Imp.$X_2^{***}$: imputation methods that respectively use $X_2^*$, $X_2^{**}$ and $X_2^{***}$ to impute missing $X_2$ in $s_1 - s_2$.

Table 2 shows the simulation results of $\boldsymbol{\beta}$ under the Type I two-phase design with $\rho_{X_{1,1}X_2} = 0.09$ and $f_2 = 1/3$. The imputation method using the correctly specified imputation model (Imp.$X_2^*$) performs similarly as Direct.$s_1$. They yield approximately unbiased estimators and the smallest variances among all the other methods considered. However, the results of Imp.$X_2^{**}$ and Imp.$X_2^{***}$ show that the imputation method induces substantial biases in log-OR's of $X_{1,2}$ (i.e., $\beta_2$) and the interaction between $X_2$ and $X_{1,2}$ (i.e., $\beta_{2,2}$) due to the misspecified imputation models for $X_2$, with the biases increase when the correlation between $X_2$ and the imputed variable (i.e., $X_2^{**}$ and $X_2^{***}$) decreases. In contrast, the calibration method does not induce biases no matter which auxiliary variable is used to calibrate the sample weights in $s_2$. The efficiency gain of the weight calibration estimators, however, depends on the choice of auxiliary variables. The proposed calibration method using score functions generally improves efficiency of all $\boldsymbol{\beta}$ esitmates. When the prediction model of $X_2$ is correctly specified, the calibration method (Calib.$X_2^*$) performs similarly as the imputation method Imp.$X_2^*$. When prediction models of $X_2$ is misspecified (Calib.$X_2^{**}$ and Calib.$X_2^{***}$), the efficiency gain decreases especially for estimating $\beta_2$ and $\beta_{2,2}$. The results of Calib.$X_2^{***}$ indicate that even though the correlation between $X_2$ and its prediction is low ($\rho_{X_2,X_2^{***}} = 0.3$), the proposed calibration method still can greatly improve the efficiency of the $\boldsymbol{\beta}$ estimates for other covariates without inducing biases. As a result, the proposed calibration method is more efficient than the traditional calibration on covariates, and more robust to a using misspecified prediction model for $X_2$ compared to the imputation method, and therefore, yields $\boldsymbol{\beta}$ estimates with smaller mean squared errors (MSE) than the imputation method in general. The proposed TL variance estimation performs well, with the variance ratios close to 1. The results under the Type II two-phase design show similar patterns (Table D.1 in the Supplementary Materials).

Figure 2 shows that the efficiency gain of the calibration method can increase when $f_2$, (i.e., the sample fraction of $s_2$ among $s_1$) decreases under both designs. The sample sizes of $s_1$ and $s_2$, i.e., $n_1$ and $n_2$, are respectively the same under the two types of two-phase designs in Figures 2(a) and 2(b) as $f_2$ varies. This allows for comparing how the proposed calibration method performs under the two types of designs with varying $f_2$. The efficiency gain increases faster when the correlation between the $X_2$ and its prediction is higher for both $\hat{\beta}_{1,2}$ (i.e., log-OR for $X_{1,2}$ with no missing values) and $\hat{\beta}_2$ (i.e., log-OR for $X_2$) under both designs. However, the efficiency gain increases slower for $\hat{\beta}_2$ than that for $\hat{\beta}_{1,2}$ as $f_2$ decreases. The patterns of efficiency gain for $\hat{\beta}_0$ and $\hat{\beta}_{1,1}$ are similar to $\hat{\beta}_{1,2}$, while the patterns of efficiency gains for $\hat{\beta}_{2,2}$ is similar to $\hat{\beta}_2$ (not shown in the paper). It is also important to notice that the efficiency gains of $\hat{\beta}_{1,2}$ and $\hat{\beta}_2$ increase faster under the Type II two-phase design than those under the Type I two-phase design. This is because, by treating samples from different survey cycles as separate strata and PSUs, the degrees of freedom for the variance (i.e., total number of PSU's−total number of strata in the sample) increases dramatically under the Type II two-phase design. Under the Type I two-phase design, however, degrees of freedom for the variance do not change due to the same number of strata and clusters in $s_1$ and $s_2$. This result suggests that, when the two-phase samples are from continuous cycles of samples, combining $s_2$ from multiple cycles can provide more efficient estimates than calibrating $s_2$ to $s_1$ in one cycle. Nevertheless, combining multiple cycles require additional assumptions as discussed in Section 2.

As it is shown in Figure 3, the high correlation between covariate of interest $X_2$ and another covariate $X_{1,1}$ available in $s_1$, $\rho_{X_{1,1}X_2}$, increases the efficiency gain of $\hat{\beta}_2$ only if the $X_2$ is poorly correlated with its prediction. The efficiency gain of $\hat{\beta}_2$ and $\hat{\beta}_{1,1}$ decreases when $\rho_{X_{1,1}X_2}$ increases. In practice, neither the true outcome model for $Y$ nor the prediction model for $X_2$ is known. A question of practical interest is whether covariates that are highly correlated with $X_2$ should be included in the outcome model for $Y$ or the prediction model for $X_2$ when the calibration method is applied. Our simulation results suggest these covariates should be included in the prediction model for $X_2$ rather than in the outcome model efficiency improvement. However, for a particular study there may be substantive reasons to include these covariates in the outcome model. Including these covariates in both outcome model and the prediction model for $X_2$ only limits the efficiency of the calibration method but does not induce biases.

## 5 DATA APPLICATION: The U.S. National Health and Nutrition Examination Survey

The U.S. National Health and Nutrition Examination Survey (NHANES) has been conducted by the National Center for Health Statistics (NCHS), Centers for Disease Control and Prevention (CDC) since 1999 is a continuously repeated biannual cross-sectional complex, multistage sample survey that represents the civilian, noninstitutionalized population of the US over the age of 2 months (NHANES 2018). NHANES combines in-person interviews that gather respondents' self-reported demographic, socioeconomic, dietary, and health-related characteristics, with physical examinations that obtain medical measurements and results of laboratory tests from biospecimens such as blood and urine. Furthermore, NHANES has been linked to the National Death Index to provide mortality follow-up data from the date of survey participation through the end of 2019 (NCHS, 2022), which serves as a nationally representative data source for both cross-sectional analyses for common diseases and health conditions and as a nationally representative cohort with all-cause and cause-specific mortality outcomes.

Many laboratory tests are only conducted among subsamples in each year (Type I Two-Phase Design), or only conducted in certain survey years (Type II Two-Phase Design). We apply the proposed calibration method to estimate the association between perspective 4-year all-cause mortality and glucose tolerance, as well as the association between assessment of diabetes and non-alcoholic fatty liver disease (NAFLD) under the Type I and II two-phase designs, respectively.

## 5.1 Association between All-Cause Mortality and Glucose Tolerance

The 2-hour fasting plasma glucose (2-HR FPG) level after the 75g oral glucose tolerance test (OGTT) serves as a gold-standard test for diagnosing diabetes and intermediate hyperglycemia (WHO, 2006). It has also been demonstrated to have a significant association with all-cause mortality (Barr et al., 2007; Inoue et al., 2021). We estimate the association between the 2-HR FPG (100 mg/dL) and prospective 4-year all-cause mortality among adults ages 18 and older, who are not diagnosed with diabetes and not pregnant in the US using $n_1 = 14,073$ NHANES sample individuals in three survey cycles: 2011-2012, 2013-2014, and 2015-2016.

The 2-HR FPG is only collected in a subsample in each NHANES cycle due to the fasting requirement and the health condition of the participants. Individuals in the full (i.e., first-phase) NHANES sample $s_1$ are randomly assigned to the morning session, or in the afternoon /evening. Only participants assigned to the morning session are required to fast overnight and provide a fasting blood sample in the examination component of NHANES. Individuals who did not report diagnosed diabetes or pregnancy and had fasted at least 9 hours were eligible for the OGTT (NHANES, 2018). The OGTT subsample $s_2$ includes $n_2 = 5,587$ individuals in $s_1$, and has the subsample weights adjusted for eligibility, fasting status, and missing values. The random serum glucose (RSG) test, which does not require fasting is a simpler single test to assess glucose levels in the blood, and it is collected in the entire $s_1$. This test is less accurate for diabetes diagnostics, and less associated with comorbidities compared to the 2-HR FPG test (Coutinho et al., 1999). However, the RSG level can be used as predictor of the 2-HR FPG level because of the high correlation between the two measurements ($\rho = 0.68$ in $s_2$).

The outcome model of all-cause mortality, besides the 2-HR FPG level (mg/dL), includes other covariates known to impact risk of all-cause mortality: self-reported demographic variables (age, sex, race/ethnicity), smoking status, and the intensity of physical activities (PA) measured by Metabolic Equivalents (World Health Organization, 2005) obtained from the interview, and the Body Mass Index (BMI) obtained from the physical examination.

The log-OR's, i.e., $\boldsymbol{\beta}$, are estimated by three methods : (1) direct $s_2$ sample weighted estimation; (2) the imputation method using RSG level to impute 2-HR FPG level in $s_1 - s_2$; and (3) the proposed weight calibration method using RSG level as the predictor of 2-HR FPG level to compute the auxiliary variables of score functions for calibration (Table 3). The direct $s_2$ sample weighted estimates of $\boldsymbol{\beta}$ differ from the imputation or calibration estimates for racial/ethnical group of Non-Hispanic Other and current smokers. This is likely due to small sample bias and the larger variances of the estimated $\boldsymbol{\beta}$ because of limited number of mortality cases in these groups in $s_2$ (43 and 124 for Non-Hispanic Other and current smokers respectively). The proposed calibration estimates of $\beta$'s, including $\beta$ for the 2-HR FPG level, are much more efficient the direct $s_2$ estimates, resulting in significant effects of Hispanic racial/ethnic group, former smokers, and intensity of PA, which the direct $s_2$ estimates fail to show. The imputation estimates are also much more efficient than the direct $s_2$ estimates, but the imputation estimate of $\beta$ for the 2-HR FPG is quite different from the direct $s_2$ and the calibration estimate, which may be caused by misspecification of the imputation model for the 2-HR FPG.

## 5.2 Association between Diabetes and Non-Alcoholic Fatty Liver Disease

Non-Alcoholic Fatty Liver Disease (NAFLD) has become the most common liver disease in Western countries and has been shown to be strongly associated with incidence of diabetes (Balkau et al., 2010; Mantovani et al., 2018). The controlled attenuation parameter (CAP) measured by a liver ultrasound transient elastography instrument called FibroScan is one of the most accurate scores to determine both the presence and the stage of hepatic steatosis (fat in the liver). It is often used as a reference to evaluate the performance of other less direct NAFLD scores (Kozłowska-Petriczko et al., 2021; Xavier et al., 2021). We estimate the association between CAP scores and presence of diabetes among non-heavy drinking adults (individuals aged 18 years or older who consume fewer than 1 drink per day for females and fewer than 2 drinks per day for males) using data from 4 NHANES cycles conducted between 2011 and 2018.

The CAP was measured among participants older than 12 years in NHANES, but only in one survey full cycle in NHANES 2017-2018 before the pandemic. In contrast, the fatty liver index (FLI) is a cheaper and easier obtained score for NFALD using biochemical and clinical examinations, and it has been shown to be highly correlated with CAP (Kozłowska-Petriczko et al., 2021). The FLI, as shown in formula (5.1), is a function of Triglycerides (TG) (mg/dL), Body Mass Index (BMI) (kg/m$^2$), gamma-glutamyl-transferase (GGT) (U/L), and waist circumference (WC) (cm) (Bedogni et al., 2006), which are collected among participants aged 2 years or older from the examinations of the in each NHANES 2-year cycle. Therefore, FLI serves as a good predictor of CAP to compute auxiliary variables of score functions due to the availability in the full combined sample (i.e., the "first-phase" sample $s_1$) and the high correlation ($\rho = 0.63$ in the NHANES 2017-2018, i.e., the "second-phase" sample $s_2$).

$$\text{FLI} = \text{expit}\{-15.745 + 0.953\log(\text{TG}) + 0.139\text{BMI} + 0.718\log(\text{GGT}) + 0.053\text{WC}\} \quad (5.1)$$

The presence of diabetes is defined by either self-reported diabetes or the following results of at least one of two laboratory tests from the physical examination: (1) blood glycohemoglobin (HbA1c) higher than 6.5%, or (2) fasting glucose higher than 126 mg/dL (ADA, 2010). The outcome model includes self-reported age, sex, race/ethnicity, intensity of PA, which are from the questionnaires, BMI from the physical examination, presence of hypertension, which is defined as having self-reported hypertension or blood pressure measured in the physical examination (a systolic pressure ≥ 130 or a diastolic pressure ≥ 80), CAP, and the interaction between CAP and sex. The full 4-cycle "first-phase" sample $s_1$ and the complete 1-cycle ("second-phase") sample $s_2$ respectively include $n_1 = 18,772$ and $n_2 = 4,386$ individuals.

Table 4 displays the log-OR's, i.e., $\boldsymbol{\beta}$ estimated using three methods: (1) the weighted $s_2$; (2) imputation method using FLI to impute CAP for $s_1 - s_2$; and (3) the calibration method using FLI to predict CAP. The calibration method dramatically reduces the variance of $\widehat{\boldsymbol{\beta}}$ for all risk factors. The efficiency of $\hat{\beta}$ for CAP and the interaction between CAP and sex is respectively improved by $0.05/0.028 = 1.79$ times and $0.112/0.073 = 1.53$ times. Besides, the calibration method shows that individuals in the overweight category have significantly higher risks of having diabetes compared to underweighted or normal weighted individuals, which is not captured by the direct $s_2$ estimator due to the high variance. The imputation method, though are even more efficient than the calibration method, yields quite different $\beta$ estimates for BMI and the main effect of CAP.

This is potentially due to BMI being a variable used in both the computation of the FLI score used to predict CAP for imputation model and as a covariate in the outcome model. This will tend to reduce the association of BMI and the presence of diabetes in the outcome model with imputed CAP as a covariate in this model. The calibration method, on the contrary, does not substantially change the $\beta$ estimates for main effect of BMI and CAP compared with the direct $s_2$ estimates.

## 6 DISCUSSION

Two-phase designs are common in epidemiologic studies and large-scale health surveys, which measure some covariates of interest in the second-phase sample only. Limiting the analysis to the complete second-phase sample can be inefficient due to reduced sample size compared to the first-phase sample. Traditional calibration weighting methods that use covariates as the auxiliary variables cannot improve efficiency of the regression analyses due to the low correlation between the covariates and coefficient estimates. This paper proposes to calibrate the second-phase sample weights to the weighted first-phase sample on score functions under the general two-phase designs, which are commonly applied in population-based epidemiologic studies and national health surveys. In simulations, the proposed weight calibration method is shown to greatly improve the efficiency in estimating regression coefficients compared to the direct second-phase sample analysis or the traditional calibration weighting methods. Moreover, different from the imputation estimators, which can be biased under a misspecified imputation model for the second-phase covariate of interest, the proposed calibration estimators are design consistent no matter the prediction model for this second-phase covariate correctly specified or not. Nevertheless, a misspecified prediction model for the second-phase covariate can reduce the correlation between this covariate and the predicted variable and therefore result in low efficiency gain by the proposed calibration method.

In the data application of risk estimation for all-cause mortality, the proposed calibration method greatly improves the efficiency in estimating the log-OR of glucose tolerance that is only measured in the second phase fasting sample, and shows significant effects of Non-Hispanic Black, former smokers, and intensity of PA, which are not captured by the complete second-phase analysis. In the application of risk estimation for diabetes, the calibration method also improves efficiency of log-OR estimation for the interaction between the second-phase covariate CAP and the first-phase variable sex.

Notice that under the type II design, assumption (A.3) applies for some surveys, including NHANES, where PSUs in the samples from different survey cycles are assumed independent (NHANES 2018), whereas in some surveys, such as the NHIS, the sampled PSUs remain the same for multiple survey cycles of the survey, but the sample households from these PSUs are different across multiple survey cycles. The variance estimation can be easily modified to account for the duplicated PSUs in combine the samples from multiple cycles of the survey.

Although this paper focuses on the logistic regression analysis, the proposed calibration method can be easily extended to other types of cross-sectional or time-to-event analyses, e.g., multiple linear regression or Cox proportional hazard regression, under two-phase survey designs. Throughout this paper, it is assumed that there is one second-phase covariate of interest, but the method can be applied to situations with second-phase outcomes, or multiple second-phase covariates of interest. Moreover, the proposed calibration method can also be applied to many other scenarios including but not limited to handling missing data due to nonresponse (Beaumont, 2005), and integrating data from different surveys, such as NHANES and NHIS, where some variables are collected in one survey but not another.

The proposed method also has limitations. First, the calibration method assumes that the weighted first- and second-phase samples represent the same target finite population. However, this assumption can be violated under both types of two-phase designs in practice, potentially resulting in bias. For instance, in the data application under the Type I design in NHANES, the RSG level (i.e., the prediction for the second-phase covariate of interest) can be influenced by the fasting status. As a result, the distribution of RSG level in the weighted fasting sample can differ from that in the weighted full NHANES sample. Under the Type II design, the target population may undergo changes over time, resulting in varying variable distributions in the weighted samples across different survey cycles. Second, the efficiency gain achieved by the calibration method can be relatively small compared to the imputation method especially when the correlation between the second-phase covariate of interest and its predictor is low.

**ACKNOWLEDGMENTS**

This work is partially supported by the Intramural Research Program of the US National Institutes of Health/National Cancer Institute (NIH/NCI). The author is grateful to Dr. Barry Graubard from NIH/NCI, and Dr. Yan Li from University of Maryland, College Park, for their very constructive comments.

Figure 1 Diagram of the proposed procedure: Step 1 creates $X_2^*$ (prediction of $X_2$) for the entire $s_1$ using the 1$^{st}$-phase covariates $\boldsymbol{X}_1$ and ancillary variables $\boldsymbol{Z}$; Step 2 computes score functions $\{\boldsymbol{u}_i^*\{\widehat{\boldsymbol{\beta}}^*\}, i \in s_1\}$ from a regression model of $Y$ on $(\boldsymbol{X}_1^T, X_2^*)^T$ fitted in $s_1$; Step 3 calibrates the 2$^{nd}$-phase sample weights $\{w_i, i \in s_2\}$ to the 1$^{st}$-phase sample weights $\{w_{1,i}, i \in s_1\}$ using $\{\boldsymbol{u}_i^*\{\widehat{\boldsymbol{\beta}}^*\}, i \in s_1\}$ as the calibration auxiliary variables

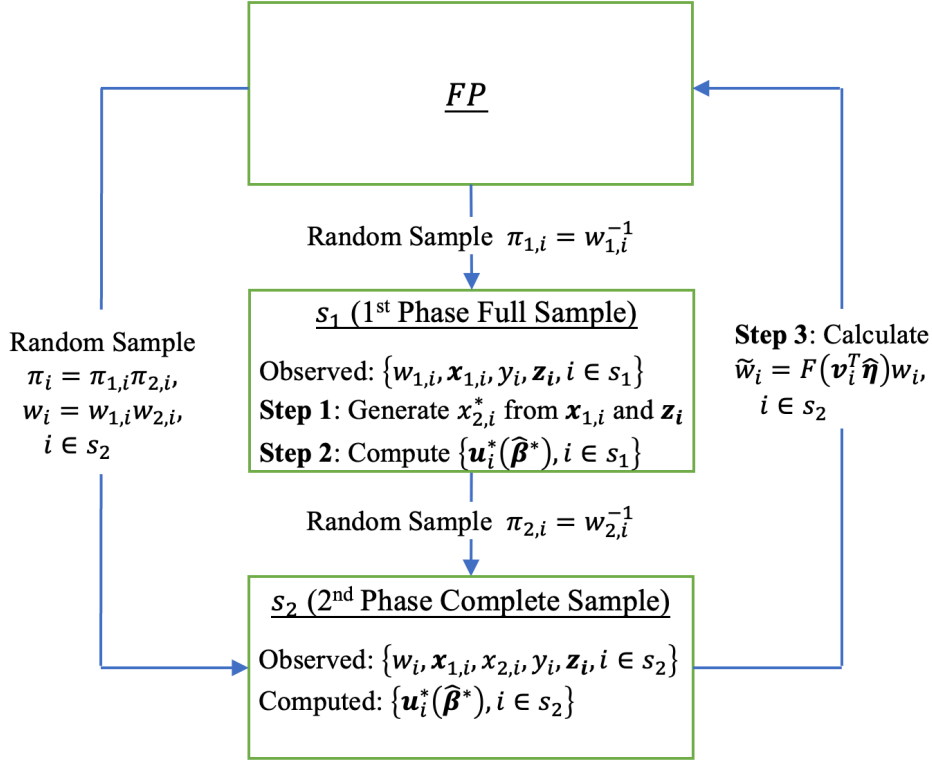

Figure 2 Efficiency gain of the calibration estimators of $\beta_{1,2}$ and $\beta_2$ compared to the direct complete sample estimators with varied fraction of complete sample out of the full sample (a) under the Type I two-phase design I and (b) under the Type II two-phase design.

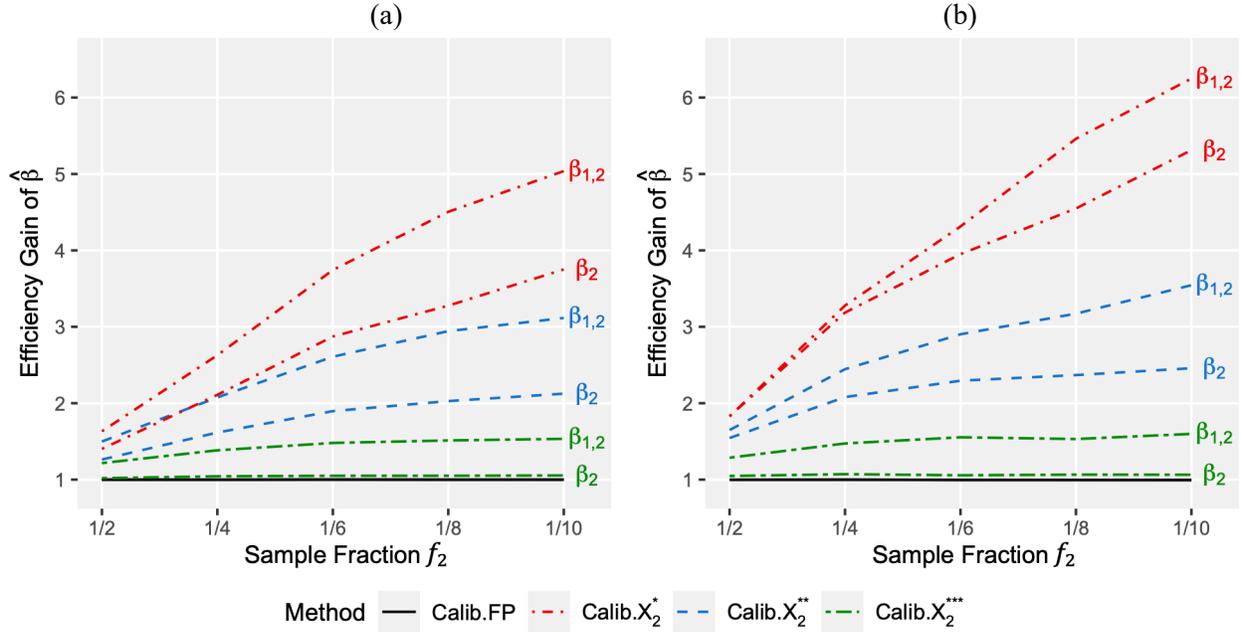

Figure 3 Efficiency gain of the calibration estimators of (a) $\beta_{1,1}$ and (b) $\beta_2$ compared to the direct 2$^{nd}$ phase sample estimators with varied $\rho_{X_{1,1}X_2}$ (correlation between $X_{1,1}$ and $X_2$) under the typical two-phase (type I) design

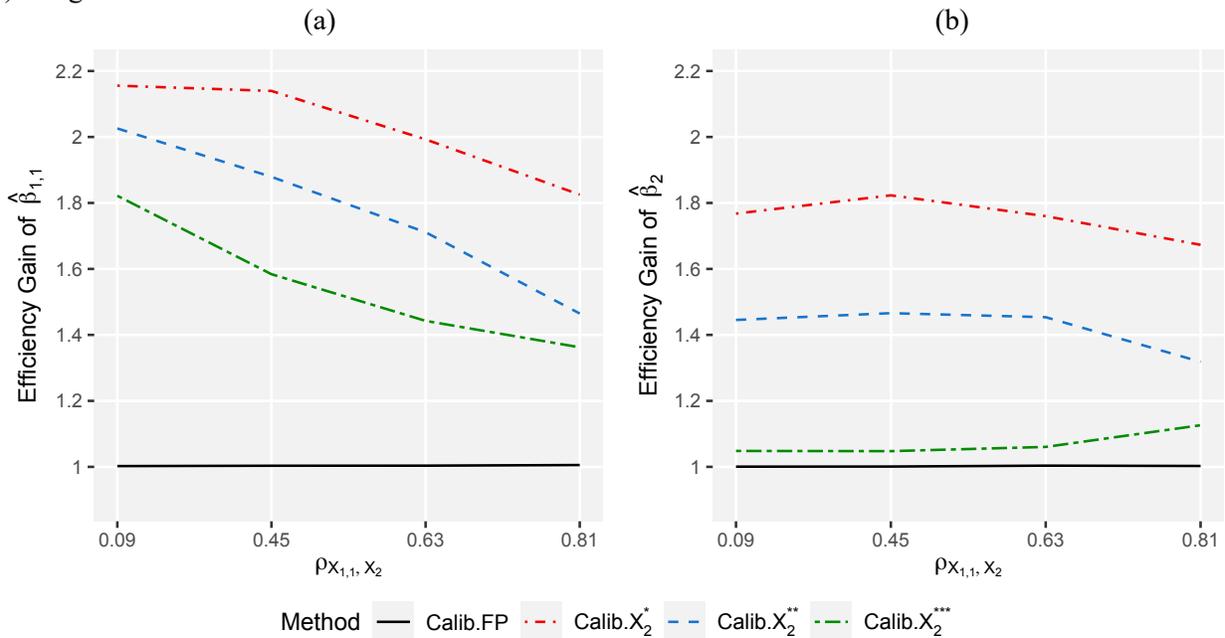

Table 1 Observational and missing scheme of the survey samples under the two types of two-phase designs

| Two-Phase Designs | Sample Weights | Outcome | Outcome Predictors | | Ancillary Variables |
| --- | --- | --- | --- | --- | --- |
| | | | Important Exposure | Other Predictors | |
| **Type I two-phase design** | | | | | |
|   Second-phase subsample | √ | √ | √ | √ | √ |
|   First-phase full sample | √ | √ | X | √ | √ |
| **Type II two-phase design** | | | | | |
|   Sample of a special cycle (second phase) | √ | √ | √ | √ | √ |
|   Combined sample of multiple cycles (first phase) | √ | √ | X | √ | √ |

Table 2 Simulation results of log-odds-ratios for $Y$ under the Type I two-phase complex sample design

| | Relative Bias (%) | | | | | Empirical Variance ($\times 10^{-3}$) | | | | | Variance Ratio (analytical/empirical) | | | | | Mean Squared Error ($\times 10^{-3}$) | | | | |
|---|---|---|---|---|---|---|---|---|---|---|---|---|---|---|---|---|---|---|---|---|
| | $\beta_0$ | $\beta_{1.1}$ | $\beta_{1,2}$ | $\beta_2$ | $\beta_{2,2}$ | $\beta_0$ | $\beta_{1.1}$ | $\beta_{1,2}$ | $\beta_2$ | $\beta_{2,2}$ | $\beta_0$ | $\beta_{1.1}$ | $\beta_{1,2}$ | $\beta_2$ | $\beta_{2,2}$ | $\beta_0$ | $\beta_{1.1}$ | $\beta_{1,2}$ | $\beta_2$ | $\beta_{2,2}$ |
| Direct.$s_2$ | 0.50 | 0.65 | -0.99 | 0.36 | 2.29 | 30.80 | 4.38 | 34.16 | 3.03 | 6.25 | 0.99 | 0.98 | 0.99 | 1.00 | 0.99 | 31.02 | 4.40 | 34.24 | 3.03 | 6.30 |
| Calib.FP | 0.50 | 0.66 | -0.97 | 0.37 | 2.26 | 30.78 | 4.37 | 34.19 | 3.03 | 6.25 | 0.99 | 0.97 | 0.99 | 0.99 | 0.99 | 31.00 | 4.39 | 34.27 | 3.03 | 6.29 |
| Calib.$X_2^*$ | 0.37 | 0.40 | -1.24 | 0.16 | 2.16 | 22.00 | 2.03 | 15.51 | 1.71 | 2.83 | 1.00 | 1.01 | 1.00 | 0.99 | 1.00 | 22.12 | 2.04 | 15.63 | 1.71 | 2.87 |
| Calib.$X_2^{**}$ | 0.40 | 0.39 | -1.22 | 0.26 | 2.20 | 23.51 | 2.16 | 18.25 | 2.09 | 3.79 | 0.99 | 1.00 | 1.00 | 0.98 | 0.99 | 23.66 | 2.17 | 18.37 | 2.10 | 3.83 |
| Calib.$X_2^{***}$ | 0.47 | 0.47 | -1.10 | 0.44 | 2.12 | 26.48 | 2.40 | 25.45 | 2.89 | 6.01 | 1.00 | 1.00 | 1.00 | 1.00 | 0.99 | 26.68 | 2.42 | 25.55 | 2.89 | 6.05 |
| Direct.$s_1$ | 0.35 | 0.40 | -1.24 | 0.13 | 2.10 | 21.26 | 1.97 | 13.90 | 1.53 | 2.28 | 0.98 | 0.99 | 0.98 | 0.97 | 0.98 | 21.37 | 1.98 | 14.03 | 1.53 | 2.32 |
| Imp.$X_2^*$ | -0.73 | -1.21 | 1.70 | -0.71 | -3.58 | 20.40 | 1.88 | 13.62 | 1.58 | 2.32 | 0.98 | 0.99 | 0.98 | 0.96 | 0.98 | 20.88 | 1.95 | 13.86 | 1.60 | 2.44 |
| Imp.$X_2^{**}$ | -2.97 | -7.79 | 5.65 | -2.10 | -16.19 | 19.43 | 1.65 | 13.91 | 1.94 | 2.74 | 0.97 | 1.00 | 0.96 | 0.94 | 0.97 | 27.38 | 4.62 | 16.50 | 2.06 | 5.10 |
| Imp.$X_2^{***}$ | -9.34 | -24.67 | 16.03 | -6.24 | -41.65 | 18.86 | 1.21 | 22.19 | 7.22 | 13.85 | 1.00 | 1.00 | 1.00 | 0.97 | 1.01 | 97.38 | 31.04 | 42.99 | 8.19 | 29.46 |

Table 3 Log-odds-ratios of 2-hour fasting plasma glucose (2-HR FPG) for 4-year all-cause mortality among adults who are not diagnosed with diabetes in the US, estimated from the second-phase Oral Glucose Tolerance Test (OGTT) sample, imputed full first-phase sample, and calibrated second-phase OGTT sample

| | Direct OGTT Sample | | | Imputation | | | Calibration | | |
|---|---|---|---|---|---|---|---|---|---|
| | $\hat{\beta}$ ($10^{-1}$) | $var(\hat{\beta})$ ($10^{-2}$) | 95% CI | $\hat{\beta}$ ($10^{-1}$) | $var(\hat{\beta})$ ($10^{-2}$) | 95% CI | $\hat{\beta}$ ($10^{-1}$) | $var(\hat{\beta})$ ($10^{-2}$) | 95% CI |
| (Intercept) | -90.92 | 45.68 | (-104.20, -77.68) | -88.07 | 18.34 | (-96.50, -79.7) | -88.88 | 15.92 | (-96.70, -81.10) |
| **Age** (in 10 years) | 7.89 | 0.89 | (6.04, 9.74) | 8.59 | 0.34 | (7.46, 9.73) | 8.19 | 0.28 | (7.16, 9.23) |
| **Sex** (Male) | 4.94 | 4.29 | (0.88, 9.00) | 4.26 | 1.03 | (2.27, 6.24) | 4.31 | 0.91 | (2.44, 6.17) |
| **Race/Ethnicity** (ref: Non-Hispanic White) | | | | | | | | | |
| Non-Hispanic Black | 2.63 | 5.21 | (-1.84, 7.11) | 2.13 | 1.63 | (-0.38, 4.63) | 1.80 | 1.41 | (-0.52, 4.13) |
| Hispanic | -2.05 | 7.52 | (-7.43, 3.32) | -2.47 | 2.74 | (-5.72, 0.78) | -3.11 | 2.50 | (-6.2, -0.01) |
| Non-Hispanic Other | -4.69 | 13.99 | (-12.02, 2.64) | -1.90 | 6.03 | (-6.71, 2.91) | -2.55 | 4.86 | (-6.87, 1.77) |
| **BMI** (ref: Underweighted or Normal) | | | | | | | | | |
| Overweight | -1.50 | 3.86 | (-5.35, 2.35) | -2.45 | 2.16 | (-5.34, 0.43) | -2.64 | 1.93 | (-5.37, 0.08) |
| Obese | -3.18 | 4.67 | (-7.42, 1.05) | -1.39 | 1.71 | (-3.95, 1.17) | -2.13 | 1.49 | (-4.53, 0.26) |
| **Smoking Status** (ref: Never Smokers) | | | | | | | | | |
| Former Smokers | 4.22 | 6.62 | (-0.82, 9.26) | 3.01 | 1.62 | (0.51, 5.50) | 2.90 | 1.48 | (0.51, 5.28) |
| Current Smokers | 12.59 | 5.91 | (7.83, 17.36) | 9.31 | 2.64 | (6.12, 12.50) | 9.38 | 2.20 | (6.47, 12.29) |
| **Intensity of PA** | -2.04 | 2.47 | (-5.12, 1.04) | -2.31 | 1.06 | (-4.34, -0.29) | -2.21 | 0.80 | (-3.96, -0.46) |
| **2-HR FPG** (10 mg/dL) | 0.564 | 0.0151 | (0.323, 0.805) | 0.293 | 0.0086 | (0.111, 0.474) | 0.550 | 0.0106 | (0.348, 0.752) |

Table 4 Log-odds-ratios of controlled attenuation parameter (CAP) for diabetes among adults in the US estimated from the NHANES 2017-18 ("second-phase") sample, imputed NHANES 2011-18 (full "first-phase") samples, and calibrated NHANES 2017-18 ("second-phase") sample

| | NHANES 17-18 Direct | | | Imputation | | | Calibration | | |
|---|---|---|---|---|---|---|---|---|---|
| | $\hat{\beta}$ ($10^{-1}$) | $var(\hat{\beta})$ ($10^{-2}$) | 95% CI | $\hat{\beta}$ ($10^{-1}$) | $var(\hat{\beta})$ ($10^{-2}$) | 95% CI | $\hat{\beta}$ ($10^{-1}$) | $var(\hat{\beta})$ ($10^{-2}$) | 95% CI |
| (Intercept) | -101.94 | 83.573 | (-119.86, -84.03) | -103.47 | 18.540 | (-111.91, -95.03) | -96.41 | 30.876 | (-107.3, -85.52) |
| **Age** (in 10 years) | 5.37 | 0.352 | (4.20, 6.53) | 4.92 | 0.046 | (4.50, 5.34) | 4.96 | 0.059 | (4.48, 5.43) |
| **Sex** (Male) | 21.47 | 115.187 | (0.43, 42.50) | 20.19 | 32.363 | (9.04, 31.34) | 19.00 | 69.129 | (2.70, 35.3) |
| **Race/Ethnicity** (ref: NH-White) | | | | | | | | | |
| NH Black | 6.90 | 2.165 | (4.02, 9.78) | 6.92 | 0.474 | (5.57, 8.27) | 8.07 | 0.469 | (6.72, 9.41) |
| Hispanic | 4.80 | 2.966 | (1.43, 8.18) | 6.19 | 0.683 | (4.57, 7.81) | 6.27 | 0.788 | (4.53, 8.01) |
| NH-Other | 8.45 | 1.887 | (5.76, 11.15) | 8.34 | 0.732 | (6.66, 10.01) | 8.18 | 0.578 | (6.69, 9.67) |
| **BMI** (ref: Underweighted or Normal) | | | | | | | | | |
| Overweight | 2.06 | 4.263 | (-1.99, 6.10) | 0.58 | 1.070 | (-1.45, 2.61) | 2.30 | 1.296 | (0.07, 4.53) |
| Obese | 6.96 | 4.581 | (2.76, 11.15) | 4.00 | 1.930 | (1.28, 6.72) | 7.67 | 1.853 | (5.00, 10.34) |
| **Hypertension** | 7.98 | 4.257 | (3.94, 12.03) | 6.41 | 0.778 | (4.68, 8.13) | 6.53 | 0.841 | (4.74, 8.33) |
| **Intensity of PA** | -0.93 | 0.189 | (-1.78, -0.07) | -0.63 | 0.060 | (-1.11, -0.15) | -0.54 | 0.031 | (-0.89, -0.20) |
| **CAP**($\times$ 10) | 1.51 | 0.050 | (1.07, 1.95) | 1.78 | 0.021 | (1.50, 2.07) | 1.40 | 0.028 | (1.07, 1.73) |
| **CAP**($\times$ 10)*Sex (Male) | -0.68 | 0.112 | (-1.34, -0.03) | -0.64 | 0.037 | (-1.02, -0.26) | -0.59 | 0.073 | (-1.11, -0.06) |